\documentclass[aps,showpacs,preprintnumbers,amsmath, amssymb]{revtex4}

\usepackage{float}
\usepackage{graphics,epsfig}
\usepackage{graphicx}
\usepackage{dcolumn}
\usepackage{bm}

\begin{document}
\baselineskip=0.8 cm
\title{{\bf Holographic entanglement entropy in general holographic superconductor models}}
\author{Yan Peng$^{1}$\footnote{yanpengphy@163.com}, Qiyuan Pan$^{2}$\footnote{panqiyuan@126.com}}
\affiliation{\\$^{1}$ School of Mathematics and Computer Science, Shaanxi University of Technology, Hanzhong, Shaanxi 723000, P. R. China  \\
$^{2}$ Institute of Physics and Department of Physics, Hunan Normal
University, Changsha, Hunan 410081, P. R. China }

\vspace*{0.2cm}
\begin{abstract}
\baselineskip=0.6 cm
\begin{center}
{\bf Abstract}
\end{center}

We study the entanglement entropy of general holographic dual models
both in AdS soliton and AdS black hole backgrounds with full
backreaction. We find that the entanglement entropy is a good probe
to explore the properties of the holographic superconductors and
provides richer physics in the phase transition. We obtain the
effects of the scalar mass, model parameter and backreaction on the
entropy, and argue that the jump of the entanglement entropy may be
a quite general feature for the first order phase transition. In
strong contrast to the insulator/superconductor system, we note that
the backreaction coupled with the scalar mass can not be used to
trigger the first order phase transition if the model parameter is
below its bottom bound in the metal/superconductor system.

\end{abstract}

\pacs{11.25.Tq, 04.70.Bw, 74.20.-z}\maketitle
\newpage
\vspace*{0.2cm}

\section{Introduction}

The anti-de Sitter/conformal field theories (AdS/CFT) correspondence
provides us a holographic dual description of the strongly coupled
field theories with a weakly coupled gravitational system
\cite{Maldacena,S.S.Gubser-1,E.Witten}. In recent years, this
correspondence has been applied to study the holographic
superconductor model, which is constructed by a scalar field coupled
to a Maxwell field in an AdS black hole background \cite{S.A.
Hartnoll,C.P. Herzog,G.T. Horowitz-1}. It shows that the black hole
becomes unstable and the scalar field condensates beyond the horizon
below a critical temperature. According to AdS/CFT correspondence,
the instability of the bulk black hole is dual to the conductor and
superconductor phase transition. Recently, the holographic dual of
the insulator and superconductor system is also established in the
background of an AdS soliton \cite{T-1}. Due to the potential
applications to the condensed matter physics, these gravity duals
attracted a lot of attention and many properties have been
disclosed, see for examples \cite{G.T. Horowitz-2}-\cite{G-1}.

The instability in the holographic superconductor models usually
corresponds to the second order phase transition. It was stated in
\cite{S. Franco,S. Franco-1} that the holographic superconductor via
the St\"{u}ckelberg mechanism allows the first order phase
transition to occur when the model parameter surpasses some
threshold value. Some interesting extensions were done in \cite{Q.Y.
Pan-1,YanPanCTP} by including backreaction. It was found that the
backreaction can trigger the first order phase transition when the
St\"{u}ckelberg model parameter is below its critical threshold. It
was announced in \cite{P. Yan} that the St\"{u}ckelberg mechanism
together with the backreaction will determine the order of phase
transition when applying the St\"{u}ckelberg mechanism to the AdS
soliton spacetime. Generally speaking, there is only the second
order phase transition for different masses of the scalar field in
the probe limit \cite{Gary. T. Horowitz-4}. Since the order of phase
transition desponds on the choices of the couplings and the mass of
the scalar field is crucial to the formation of the scalar hair in
the superconductor model, it is interesting to explore the effect of
the scalar mass $m$ on the order of phase transition in the
St\"{u}ckelberg model especially with backreaction.

On the other hand, the entanglement entropy is a powerful tool to
keep track of the degrees of freedom in a strongly coupled system
when other traditional probes might not be available. According to
the AdS/CFT correspondence, the entanglement entropy may provide us
new insights into the quantum structure of the spacetime
\cite{M-1,M-2}. Ryu and Takayanagi \cite{S-1,S-2} have presented a
proposal to compute the entanglement entropy of CFTs from the
minimal area surface in the gravity side. This proposal provides a
simple and elegant way to calculate the entanglement entropy of a
strongly coupled system which has a gravity dual. Since then, there
have been a lot of works studying the entanglement entropy in
various gravity theories
\cite{NishiokaJHEP,KlebanovNPB,PakmanJHEP,NishiokaJPA,HungJHEP,BoerJHEP,OgawaJHEP,AlbashJHEP,MyersJHEP}.
Extending the investigation to the holographic superconductors,
Albash and Johnson observed in the metal/superconductor system that
the entanglement entropy in superconducting case is always less than
the one in the metal phase and the entropy as a function of
temperature is found to have a discontinuous slop at the transition
temperature $T_{c}$ in the case of the second order phase transition
\cite{T-6}. However, there is a discontinuous jump in the entropy
when including the first order phase transition \cite{T-6}, which
means that the entropy can be used to determine the order of phase
transition. More recently, Kuang \emph{et al.} examined the
properties of the entanglement entropy in the four-dimensional AdS
black hole and found that near the contact interface of the
superconducting to normal phase the entanglement entropy has a
different behavior due to the proximity effect \cite{K}. In the
insulator/superconductor transition, it is shown that the
entanglement entropy for a half space first increases and reaches
its maximum at a certain chemical potential and then decreases
monotonically as chemical potential increases
\cite{cai-3,cai-2,Weiping Yao}. As a further step along this line,
it is of great interest to generalize the investigation on the
entanglement entropy of general holographic superconductors via the
St\"{u}ckelberg mechanism and study systematically the effects of
the mass, model parameter and backreaction on the entropy.
Furthermore we want to obtain some general feature for the
entanglement entropy of the holographic dual models both in the
backgrounds of the AdS soliton and AdS black hole.

The outline of this work is as follows. In section II, we will study
the entanglement entropy of the general superconductors in the AdS
soliton. In section III, we will extend our discussion to the AdS
black hole. We will conclude our main results in the last section.

\section{General superconductor in AdS soliton}

\subsection{Bulk equations of motion and boundary conditions}

In the probe limit, it was argued that only second order phase
transition can happen in the AdS soliton background \cite{T-1}.
Considering the backreaction of the matter field to the background,
it was found that strong backreaction can bring about first order
phase transition \cite{G-1}. Applying the St\"{u}ckelberg mechanism
to the soliton configuration, it concluded that when the
backreaction of the matter field becomes weaker, the St\"{u}ckelberg
parameter combined with the backreaction can accommodate the first
order phase transition to occur \cite{P. Yan}. Applying the
St\"{u}ckelberg mechanism to insulator/superconductor phase
transition in the five-dimensional AdS soliton spacetime, it stated
in \cite{cai-2} that the entanglement entropy serves as a good probe
to the order of phase transition. We will generalize the discussion
in \cite{cai-3,cai-2,Weiping Yao} to the more general superconductor
by choosing various masses $m$ and charges $q$ of the scalar field,
and examining the formation of scalar hair in another
St\"{u}ckelberg superconductor model which is different from that in
Ref. \cite{cai-2}.

The generalized St\"{u}ckelberg Lagrange density reads \cite{P. Yan}
\begin{eqnarray}\label{lagrange-1}
\mathcal{L}=R+\frac{(d-1)(d-2)}{L^{2}}-\frac{1}{4}F^{\mu\nu}F_{\mu\nu}-(\partial
\psi)^{2}-m^{2}|\psi|^{2}-F(\psi)(\partial p-qA)^{2},
\end{eqnarray}
where $\psi(r)$ and $A_{\mu}$ are the scalar and Maxwell fields, $d$
is the dimensionality of the spacetime, and $L$ is the AdS radius
which will be scaled unity in our calculation. Here we will change
the strength of backreaction with the charge of the scalar field
$q$. When $q\rightarrow \infty$ with the fixed $q\psi$ and $q\phi$,
the backreaction of the matter fields becomes negligible and the
metric solutions reduce to the pure AdS soliton spacetime. For the
general function $F(\psi)$, in contrast to
$F(\psi)=\psi^{2}+\zeta\psi^{6}$ discussed in Ref. \cite{cai-2}, we
will choose $F(\psi)=\psi^{2}+q^{2}c_{4}\psi^{4}$ \cite{P. Yan} in
this work, where $c_{4}$ is the model parameter. Setting
$qA=\hat{A}$ and considering the gauge symmetry
$\hat{A}\rightarrow \hat{A}+\partial\Lambda$ and $p\rightarrow
p+\Lambda$, we can fix the gauge $p=0$ by using the gauge freedom.

Since we are interested in including the backreaction, we will
choose the metric in the form \cite{G-1}
\begin{eqnarray}\label{metric}
ds^{2}=r^{2}[-e^{C(r)}dt^{2}+e^{D(r)}B(r)d\eta^{2}+dx^{2}+dy^{2}]+\frac{dr^{2}}{r^{2}B(r)},
\end{eqnarray}
where we require that $B(r)$ vanishes at some radius $r_{0}$ which
is the tip of the soliton. In order to smooth the solutions at the
tip, we should impose a period $\kappa$ for the coordinate $\eta$
\begin{eqnarray}\label{period}
\kappa=\frac{4\pi e^{-D(r_{0})/2}}{r_{0}^{2}B'(r_{0})}.
\end{eqnarray}

Choosing the Maxwell and scalar fields in the form
\begin{eqnarray}\label{Fields}
A=\phi(r)dt,~~\psi=\psi(r),
\end{eqnarray}
we can obtain the equations of motion
\begin{eqnarray}\label{psi}
\psi''+\left(\frac{5}{r}+\frac{B'}{B}+\frac{C'}{2}+\frac{D'}{2}\right)\psi'+\frac{q^{2}\phi^{2}e^{-C}}{r^{4}B}\left(\psi+2q^{2}c_{4}\psi^{3}\right)-\frac{m^{2}}{r^{2}B}\psi=0,
\end{eqnarray}
\begin{eqnarray}\label{phi}
\phi''+\left(\frac{3}{r}+\frac{B'}{B}-\frac{C'}{2}+\frac{D'}{2}\right)\phi'-\frac{2q^{2}\phi}{r^{2}B}\left(\psi^{2}+q^{2}c_{4}\psi^{4}\right)=0,
\end{eqnarray}
\begin{eqnarray}\label{Cr}
C''+\frac{1}{2}C'^{2}+\left(\frac{5}{r}+\frac{B'}{B}+\frac{D'}{2}\right)C'-\left[\phi'^{2}+\frac{2q^{2}\phi^{2}}{r^{2}B}\left(\psi^{2}+q^{2}c_{4}\psi^{4}\right)\right]\frac{e^{-C}}{r^{2}}=0,
\end{eqnarray}
\begin{eqnarray}\label{Br}
B'\left(\frac{3}{r}-\frac{C'}{2}\right)+B\left(\psi'^{2}-\frac{1}{2}C'D'+\frac{e^{-C}\phi'^{2}}{2r^{2}}+\frac{12}{r^{2}}\right)+\frac{q^{2}\phi^{2}e^{-C}}{r^{4}}\left(\psi^{2}+q^{2}c_{4}\psi^{4}\right)
+\frac{1}{r^{2}}\left(m^{2}\psi^{2}-12\right)=0,
\end{eqnarray}
\begin{eqnarray}\label{Ar}
D'=\frac{2r^{2}C''+r^{2}C'^{2}+4rC'+4r^{2}\psi'^{2}-2e^{-C}\phi'^{2}}{r(6+rC')}.
\end{eqnarray}
Since the equations are coupled and nonlinear, we have to count on
the numerical approach. We will integrate these equations from the
tip $r_{0}$ out to the infinity.

At the tip, there are four independent parameters $r_{0}$,
$\psi(r_{0})$, $\phi(r_{0})$ and $C(r_{0})$. Considering the two
useful scaling symmetries
\begin{eqnarray}\label{symmetry-1}
r \rightarrow ar,~~~~~~~~(t,\eta,x,y)\rightarrow
~(t,\eta,x,y)/a,~~~~~~~\phi\rightarrow a \phi,
\end{eqnarray}
\begin{eqnarray}\label{symmetry-2}
C \rightarrow C-2\ln b,~~~~~~~~t\rightarrow ~b
t,~~~~~~~\phi\rightarrow \phi/b,
\end{eqnarray}
we can adjust the solutions to satisfy $r_{0}=1$ and $C(r_{0})=0$.
At $r\rightarrow \infty$, after choosing
$m^{2}\geqslant m_{BF}^{2}=-\frac{(d-1)^{2}}{4}=-4$ \cite{P. Breitenlohner},
the scalar and Maxwell fields have the form
\begin{eqnarray}\label{inf}
\psi=\frac{\psi_{-}}{r^{\lambda_{-}}}+\frac{\psi_{+}}{r^{\lambda_{+}}}+\cdot\cdot\cdot,\
\phi=\mu-\frac{\rho}{r^{2}}+\cdot\cdot\cdot, \ \
\end{eqnarray}
where $\lambda_{\pm}=2\pm\sqrt{4+m^{2}}$ are the conformal
dimensions of the operators, $\mu$ and $\rho$ can be interpreted as
the chemical potential and charge density in the dual theory
respectively. We will fix $\psi_{-}=0$ and use $\psi_{+}=<O_{+}>$ to
describe the phase transition in the following discussion. In order
to recover the pure AdS boundary, we also need
$C(r\rightarrow\infty)=0$ and $D(r\rightarrow\infty)=0$. It should
be noted that, after obtaining the solutions, we will scale them to
satisfy $\kappa=\pi$ \cite{G-1}.

\subsection{Holographic entanglement entropy in insulator/superconductor transition}

In this section, we want to explore the properties of the phase
transition through the topological entanglement entropy method. The
authors in Refs. \cite{S-1,S-2} have presented a proposal to compute
the entanglement entropy of conformal field theories (CFTs) from the
minimal area surface in gravity side. Consider a strongly coupled
field theory with gravity dual, the entanglement entropy of
subsystem $\bar{A}$ with its complement is given by searching for the
minimal area surface $\gamma_{\bar{A}}$ in the bulk with the same boundary
$\partial \bar{A}$ of a region $\bar{A}$. Then the entanglement entropy of $\bar{A}$
with its complement is given by
\begin{eqnarray}\label{EEntropy}
S_{\bar{A}}=\frac{Area(\gamma_{\bar{A}})}{4G_{N}},
\end{eqnarray}
where $G_{N}$ is the Newton's constant in the bulk. For simplicity,
we consider the entanglement entropy for a half space which
corresponds to a subsystem $\bar{A}$ defined by $x>0$,
$-\frac{R}{2}<y<\frac{R}{2}$ ($R\rightarrow\infty$),
$0\leq\eta\leq\kappa$. Then the entanglement entropy can be deduced
from Eq. (\ref{EEntropy}) as \cite{cai-3,cai-2,Weiping Yao}
\begin{eqnarray}\label{EEntropyReal}
S^{half}_{\bar{A}}=\frac{R\kappa}{4G_{N}}\int^{\frac{1}{\varepsilon}}_{r_{0}}re^{\frac{D(r)}{2}}dr=\frac{R\pi}{8G_{N}}\left(\frac{1}{\varepsilon^{2}}+S\right),
\end{eqnarray}
where $r=\frac{1}{\varepsilon}$ is the UV cutoff. The first term is
divergent as $\varepsilon\rightarrow0$. In contrast, the second term
does not depend on the cutoff and thus is physical important. As a
matter of fact, this finite term is the difference between the
entropy in the pure AdS soliton and the pure AdS space, and $S=-1$
corresponds to the pure AdS soliton.

\begin{figure}[h]
\includegraphics[width=155pt]{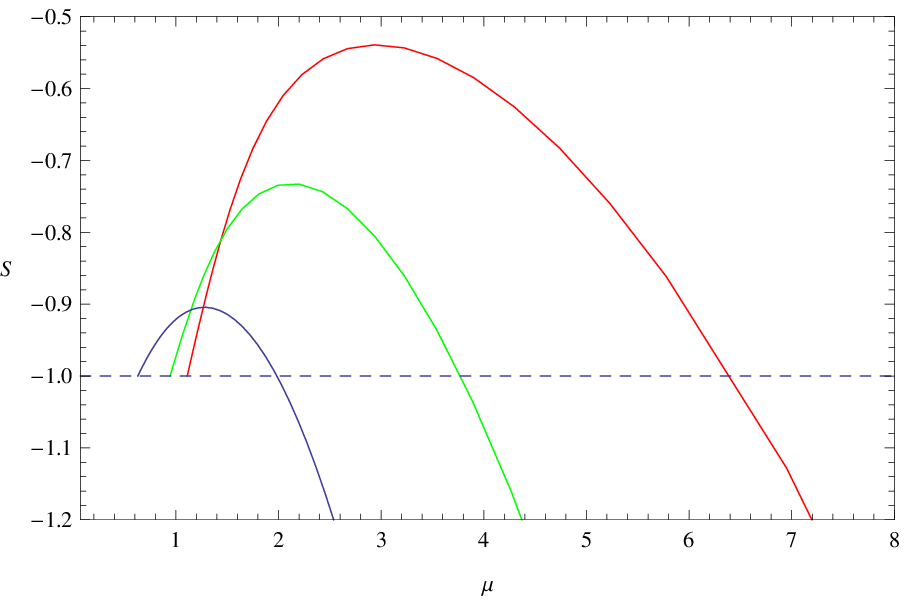}\
\includegraphics[width=155pt]{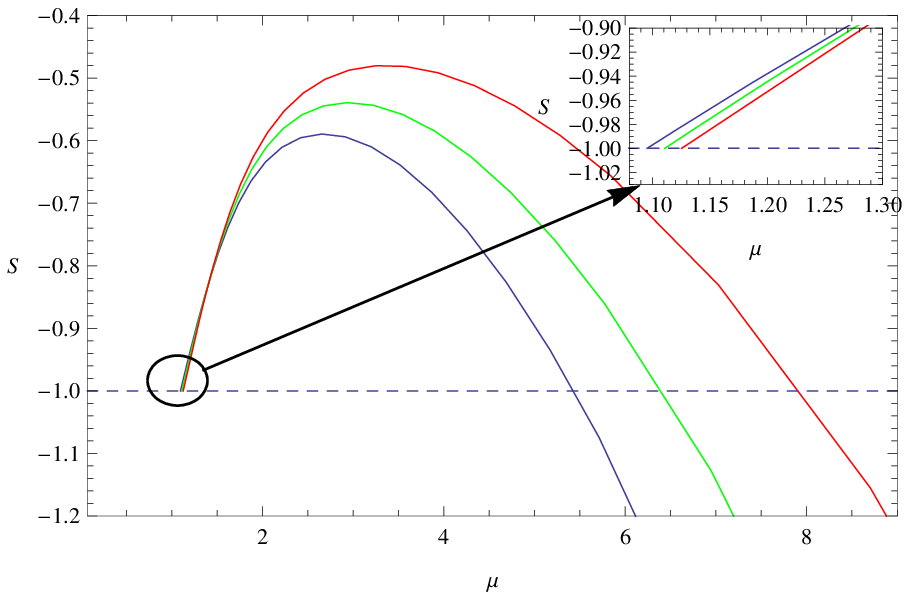}\
\includegraphics[width=155pt]{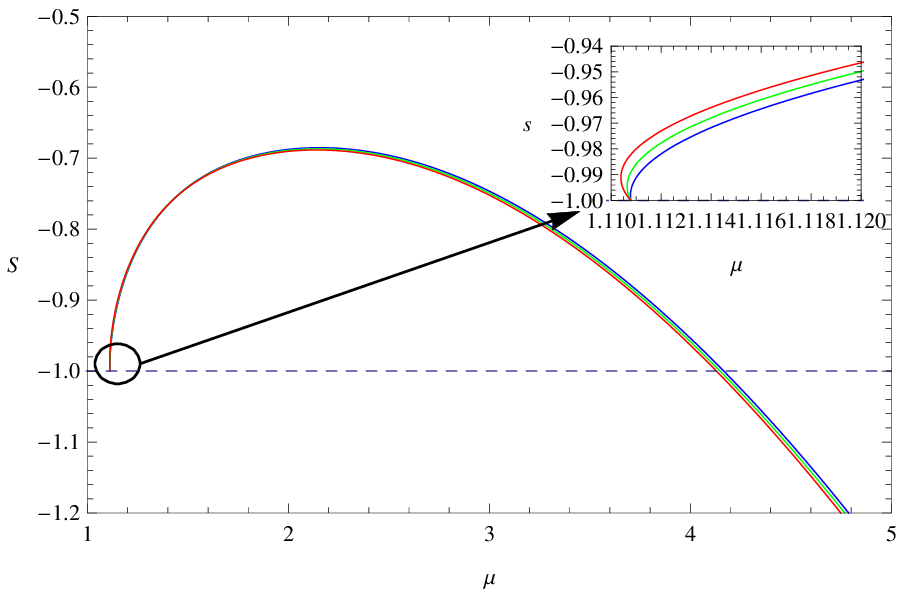}\
\caption{\label{EEntropySoliton} (Color online) The entanglement
entropy as a function of the chemical potential $\mu$ for
$\kappa=\pi$. The dashed blue line in each panel corresponds to the
entropy without backreaction or the entropy of pure AdS soliton
solution. The left panel is for the case $c_{4}=0$, $m^{2}=-15/4$
and the three lines from top to bottom correspond to increasing $q$,
i.e., $q=1.7$ (red), $q=2$ (green) and $q=3$ (blue) respectively.
The middle one shows the case $q=1.7$, $c_{4}=0$ and the three lines
from top to bottom correspond to decreasing $m^{2}$, i.e.,
$m^{2}=-149/40$ (red), $m^{2}=-15/4$ (green) and $m^{2}=-151/40$
(blue) respectively. The right one presents the case $m^{2}=-15/4$,
$q=1.7$, and the three lines correspond to
decreasing $c_{4}$, i.e., $c_{4}=0.46$ (red), $c_{4}=0.45$ (green)
and $c_{4}=0.44$ (blue) respectively. }
\end{figure}

Now we are in a position to study the effects of the charge $q$,
mass $m$ and model parameter $c_{4}$ on the entanglement entropy. In
the left panel of Fig. \ref{EEntropySoliton}, we present the value
of the entanglement entropy $S$ as a function of chemical potential
$\mu$ with $c_{4}=0$, $m^{2}=-15/4$ for different charges $q$ in the
superconductor phase. In order to compare with the result obtained
in Refs. \cite{cai-3,cai-2,Weiping Yao}, we also give the curve for
the case $q=2$. From the picture, we can see that the entropy is a
constant, i.e., $S=-1$ in the insulator phase. After condensate, the
entropy first rises and arrives at its maximum as the chemical
potential $\mu$ increases, then decreases monotonously. Obviously,
for each value of the charge $q$, there is a discontinuity in the
slope of $S$ at the critical chemical potential $\mu_{c}$, which
indicates that the second order phase transition occurs.
Furthermore, we find that the larger critical chemical potential
$\mu_{c}$ corresponds to the larger maximum of the entropy $S$ after
the scalar field condensates.

In the middle panel of Fig. \ref{EEntropySoliton}, we show the
behavior of the entanglement entropy $S$ as a function of chemical
potential $\mu$ with $q=1.7$, $c_{4}=0$ for different masses $m$.
For each value of the mass $m$, after condensate, the entropy first
rises and arrives at its maximum as the chemical potential $\mu$
increases, then decreases monotonously. Similar to the left panel,
there is a discontinuity in the slope of $S$ at the critical
chemical potential $\mu_{c}$, which can be regarded as the signature
of the second order phase transition. Again, we see that the larger
$\mu_{c}$ corresponds to the larger maximum of $S$ after condensate.

\begin{figure}[h]
\includegraphics[width=220pt]{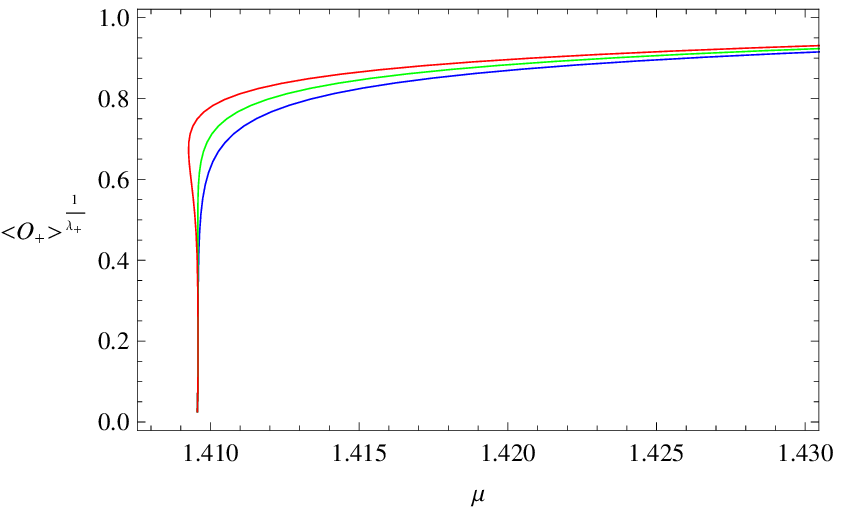}\
\includegraphics[width=220pt]{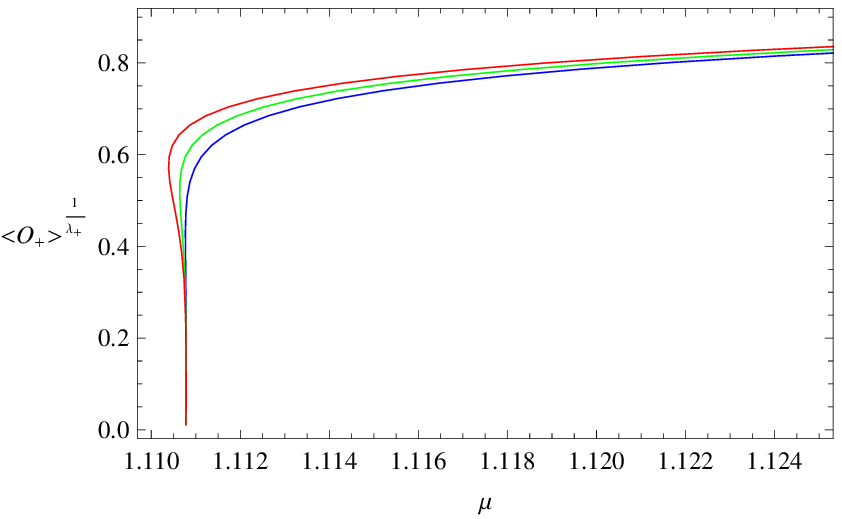}\
\caption{\label{CondensateSoliton} (Color online) The condensate
$<O_{+}>^{\frac{1}{\lambda_{+}}}$ as a function of the chemical
potential $\mu$ for $\kappa=\pi$. The three lines in the left panel
from top to bottom correspond to decreasing $c_{4}$, i.e.,
$c_{4}=0.46$ (red), $c_{4}=0.45$ (green), $c_{4}=0.44$ (blue) for
the fixed $m^{2}=-15/4$ and $q=1.7$. The three lines in the right
one correspond to $c_{4}=0.26$ (red), $c_{4}=0.25$ (green),
$c_{4}=0.24$ (blue) for the fixed $m^{2}=-3$ and $q=1.7$.}
\end{figure}

In the right panel of Fig. \ref{EEntropySoliton}, we plot the
entanglement entropy $S$ as a function of chemical potential $\mu$
with $m^{2}=-15/4$, $q=1.7$ for different model parameters $c_{4}$.
We want to examine the entropy by allowing the first order phase
transition to occur. Similar to the findings obtained in Ref.
\cite{cai-2}, we find that the entropy becomes multivalued near the
critical chemical potential $\mu_{c}$ when $c_{4}\geqslant 0.45$.
Obviously, there is a sudden jump in the entropy, which indicates a
first order phase transition there. This is in good agreement with
the results in the left panel of Fig. \ref{CondensateSoliton}, where
we exhibit the condensate of $<O_{+}>$ for selected values of the
charge $q$, mass $m$ and model parameter $c_{4}$. It should be noted
that, when neglecting the backreaction of the matter fields on the
background, the topological entropy is always a constant, i.e.,
$S=-1$ and we can not distinguish the order of phase transition.

For clarity, we also detect the effect of the mass $m$ on the
condensation in this general insulator/superconductor model, which
is missing in our previous work \cite{P. Yan}. Choosing $q=1.7$,
$m^{2}=-3$ and $-15/4$, we show the condensate
$<O_{+}>^{1/\lambda_{+}}$ as a function of the chemical potential
$\mu$ for different values of $c_{4}$ in Fig.
\ref{CondensateSoliton}. We see that there is a threshold value
$\overline{c}_{4}$ of $c_{4}$. When we enhance $c_{4}$ across the
threshold, the condensate operator does not have a monotonic
behavior, which indicates that the holographic
insulator/superconductor system in AdS soliton experiences a first
order phase transition. For the fixed $q=1.7$, we find
$\overline{c}_{4}=0.25$ and $\overline{c}_{4}=0.45$ corresponds to
the cases of $m^{2}=-3$ and $m^{2}=-15/4$ respectively, which means
that the threshold value of $c_{4}$ will decrease as the mass
$m^{2}$ increases. Thus, we conclude that for the fixed $q$ and
$c_{4}$, the more negative mass $m^{2}$ make the first order phase
transition harder to occur. Moreover, the appearance of the first
order phase transition in Fig. \ref{CondensateSoliton} can be used
to back up the numerical findings in the entanglement entropy $S$
shown in the right panel of Fig. \ref{EEntropySoliton}.

\section{General superconductor in AdS black hole}

\subsection{Bulk equations of motion and boundary conditions}

It was announced in \cite{T-6} that the belt entanglement entropy
experiences a jump when allowing the first order phase transition to
occur in the four-dimensional AdS black hole background. In this
section, we will extend the discussion by including the first order
phase transition through St\"{u}ckelberg mechanism. Taking
backreaction of the spacetime into account, we take the ansatz for
the metric of the four-dimensional AdS black hole
\begin{eqnarray}\label{AdSBH}
ds^{2}=-g(r)e^{-\chi(r)}dt^{2}+\frac{dr^{2}}{g(r)}+r^{2}(dx^{2}+dy^{2}).
\end{eqnarray}
It requires that $g(r)$ vanishes at some radius $r_{+}$ which
corresponds to the horizon of the black hole. So the Hawking
temperature reads
\begin{eqnarray}\label{HawkingT}
T_{H}=\frac{g'(r_{+})e^{-\chi(r_{+})/2}}{4\pi}.
\end{eqnarray}

Assuming the matter fields in the forms
\begin{eqnarray}\label{MatterFieldBH}
A=\phi(r)dt,~~\psi=\psi(r),
\end{eqnarray}
We can obtain equations of motion
\begin{eqnarray}\label{BHChi}
\chi'+\left[r\psi'^{2}+\frac{r}{g^{2}}e^{\chi}\phi^{2}\left(\psi^{2}+q^{2}c_{4}\psi^{4}\right)\right]=0,
\end{eqnarray}
\begin{eqnarray}\label{BHg}
g'-\left(\frac{3r}{L^{2}}-\frac{g}{r}\right)+
rg\left[\frac{1}{2}\psi'^{2}+\frac{1}{4g}e^{\chi}\phi'^{2}+\frac{m^{2}}{2g}\psi^{2}+\frac{1}{2g^{2}}e^{\chi}\phi^{2}\left(\psi^{2}+q^{2}c_{4}\psi^{4}\right)\right]=0,
\end{eqnarray}
\begin{eqnarray}\label{BHphi}
\phi''+\left(\frac{2}{r}+\frac{\chi'}{2}\right)\phi'-\frac{2\left(\psi^{2}+q^{2}c_{4}\psi^{4}\right)}{g}\phi=0,
\end{eqnarray}
\begin{eqnarray}\label{BHpsi}
\psi''+\left(\frac{2}{r}-\frac{\chi'}{2}+\frac{g'}{g}\right)\psi'-\frac{m^{2}}{g}\psi+\frac{1}{g^{2}}e^{\chi}\phi^{2}\left(\psi+2q^{2}c_{4}\psi^{3}\right)=0.
\end{eqnarray}
Using the shooting method, we can solve these equations of motion
numerically by integrating them from the horizon out to the
infinity.

At the horizon, there are four independent parameters $r_{+}$,
$\psi(r_{+})$, $\phi'(r_{+})$ and $\chi(r_{+})$. Considering the
symmetry
\begin{eqnarray}\label{symmetryBH}
r \rightarrow ar,~~~~~~~~(t,x,y)\rightarrow
~(t,x,y)/a,~~~~~~~\phi\rightarrow a
\phi,~~~~~~g\rightarrow\ a^{2} g,
\end{eqnarray}
we can adjust the solutions to satisfy $r_{+}=1$. At the asymptotic
AdS boundary $(r\rightarrow \infty)$, after choosing $m^{2}$ above
the BF bound $m^{2}\geq m_{BF}^{2}=-(d-1)^{2}/4=-9/4$ \cite{P.
Breitenlohner}, the scalar and Maxwell fields behave like
\begin{eqnarray}\label{InfBH}
\psi=\frac{\psi_{-}}{r^{\lambda_{-}}}+\frac{\psi_{+}}{r^{\lambda_{+}}}+\cdot\cdot\cdot,\
\phi=\mu-\frac{\rho}{r}+\cdot\cdot\cdot, \ \
\end{eqnarray}
with $\lambda_{\pm}=(3\pm\sqrt{9+4m^{2}})/2$. Just as in the models
of AdS soliton, we also take $\psi_{-}=0$ and the scalar
condensation is described by the operator $\psi_{+}=<O_{+}>$. After
obtaining the superconducting solutions, we will take the
transformation $q\psi\rightarrow\frac{1}{\sqrt{2}}\psi,
q\phi\rightarrow\phi$, $\frac{c_{4}}{2}\rightarrow c_{4}$, and use
$\gamma=\frac{1}{q^{2}}$ to describe the strength of backreaction
\cite{Q.Y. Pan-1,Sean A. Hartnoll-3}. Note that this transformation
does not change the topological entanglement entropy and the order
of phase transitions. When $\gamma\rightarrow 0$, i.e.,
$q\rightarrow \infty$ with the fixed $q\psi$ and $q\phi$, it reduces
to the standard holographic model in the absence of backreaction
\cite{S. Franco,S. Franco-1,Q.Y. Pan-1}.

\subsection{Holographic entanglement entropy in superconductor transition}

It was found in metal/superconductor system that the entanglement
entropy in superconducting case is always less than the one in the
metal phase and the entropy as a function of temperature is found to
have a discontinuous slop at the transition temperature $T_{c}$ in
the case of second order phase transition \cite{K,T-6}. In this
section, we want to continue the discussion by examining the effects
of the backreaction $\gamma$, mass $m$ and model parameter $c_{4}$
on the entropy.

Consider the subsystem $\tilde{A}$ with a straight strip geometry
described by $-\frac{l}{2}\leqslant x\leqslant\frac{l}{2},~0\leq y \leq \tilde{L}$,
where $l$ is defined as the size of region $\tilde{A}$ and
$\tilde{L}$ is a regulator which can be set to infinity. Minimizing
the area of hypersurface $\gamma_{\tilde{A}}$ whose boundary is the
same as the stripe $\tilde{A}$, the entanglement entropy for a belt
geometry can be expressed as \cite{T-6}
\begin{eqnarray}\label{EEntropyBH}
S=\int^{z_{*}}_{\varepsilon}dz\frac{z_{*}^{2}}{z^{2}}\frac{1}{\sqrt{(z^{4}_{*}-z^{4})z^{2}g(z)}}-\frac{1}{\varepsilon},
\end{eqnarray}
with
\begin{eqnarray}\label{Length}
\frac{l}{2}=\int^{z_{*}}_{\varepsilon}dz\frac{z^{2}}{\sqrt{(z^{4}_{*}-z^{4})z^{2}g(z)}},
\end{eqnarray}
where $z_{*}$ satisfies the condition $\frac{dz}{dx}|_{z_{*}}=0$
with $z=\frac{1}{r}$.

\begin{figure}[h]
\includegraphics[width=220pt]{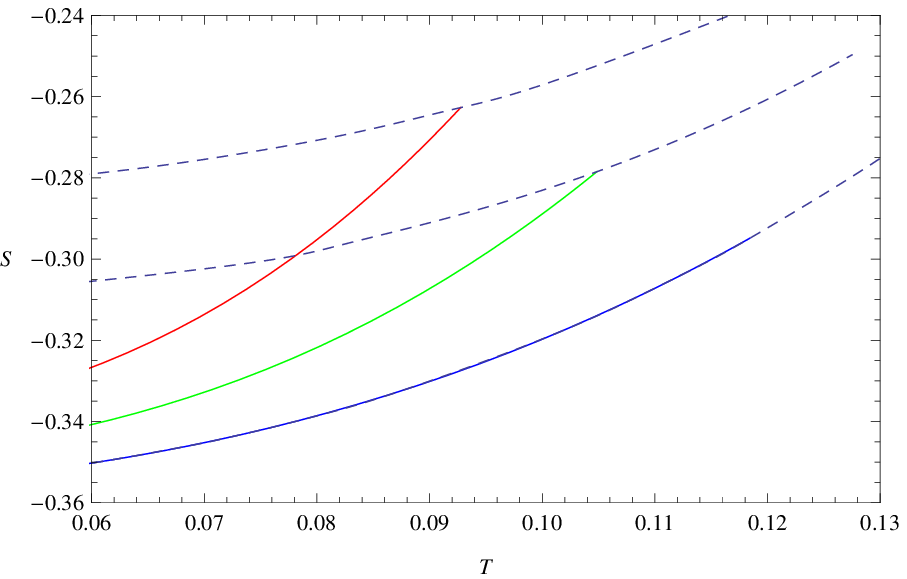}\
\includegraphics[width=220pt]{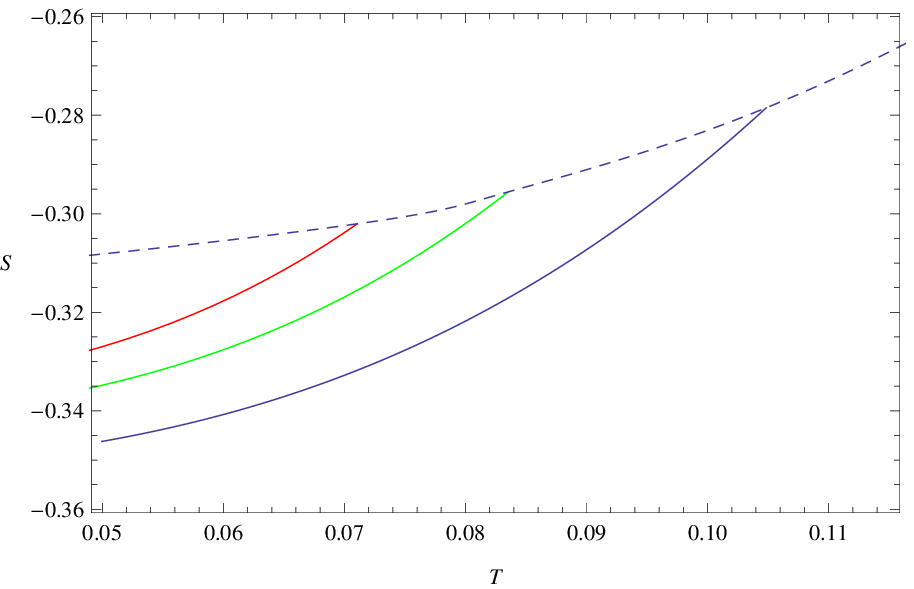}\
\caption{\label{EEntropyBH} (Color online) The entanglement entropy
as a function of temperature $T$ for fixed $l=1$ and $c_{4}=0$ with $\rho=1$. The
left panel is for the case $m^{2}=-2$, the dashed line is from the
Reissner-Nordstr\"{o}m AdS black holes and the solid curve is from
the superconductor solutions. From top to bottom, the three sets of
lines correspond to decreasing $\gamma$, i.e., $\gamma=0.2$ (red),
$\gamma=0.1$ (green) and $\gamma=0$ (blue) respectively. Similarly,
the right panel represents the case $\gamma=0.1$, and the three
solid lines from top to bottom correspond to decreasing $m^{2}$,
i.e., $m^{2}=-1$ (red), $m^{2}=-1.5$ (green) and $m^{2}=-2$ (blue)
respectively.}
\end{figure}

The entanglement entropy as a function of temperature $T$ with
different values of the backreaction $\gamma$ and mass $m$ for fixed
$l=1$ and $c_{4}=0$ is shown in Fig. \ref{EEntropyBH}. We find that,
away from the probe limit, i.e., $\gamma\neq0$, there is a
discontinuity in the slope of $S$ at the critical temperature, which
indicates the second order phase transition to occur. After
condensate, the entropy decreases monotonously, which is in
agreement with the conclusion obtained in \cite{K,T-6}. From the
picture, it also can be concluded that the critical temperature
increases if we decrease $\gamma$ or $m^{2}$. Furthermore, we can
get a relation, i.e., the higher critical temperature corresponds to
the smaller entropy. When $\gamma=0$, the topological entropy and
its slope are continuous around the critical temperature. Thus, we
can not determine the order of phase transition in the probe limit,
which is reasonable since we have neglected the backreaction of
matter fields on the metric.

\begin{figure}[h]
\includegraphics[width=220pt]{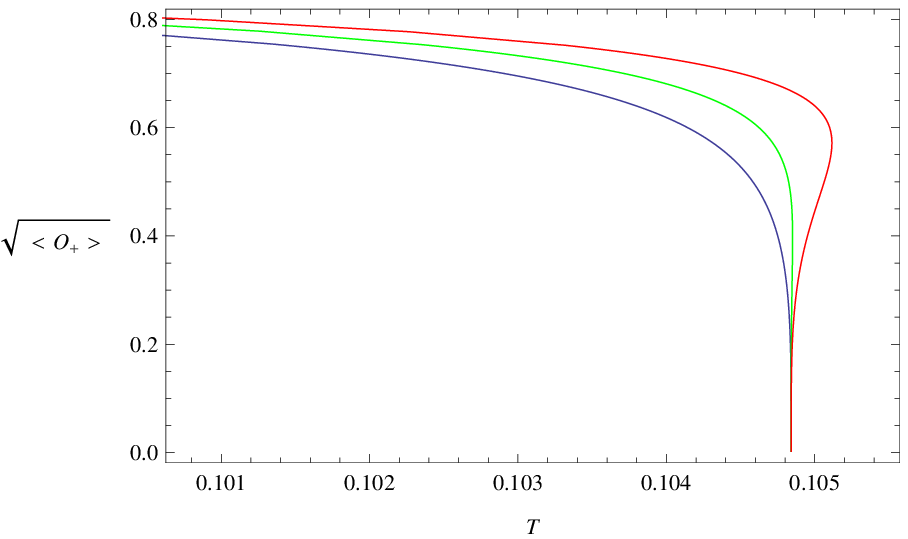}\
\includegraphics[width=220pt]{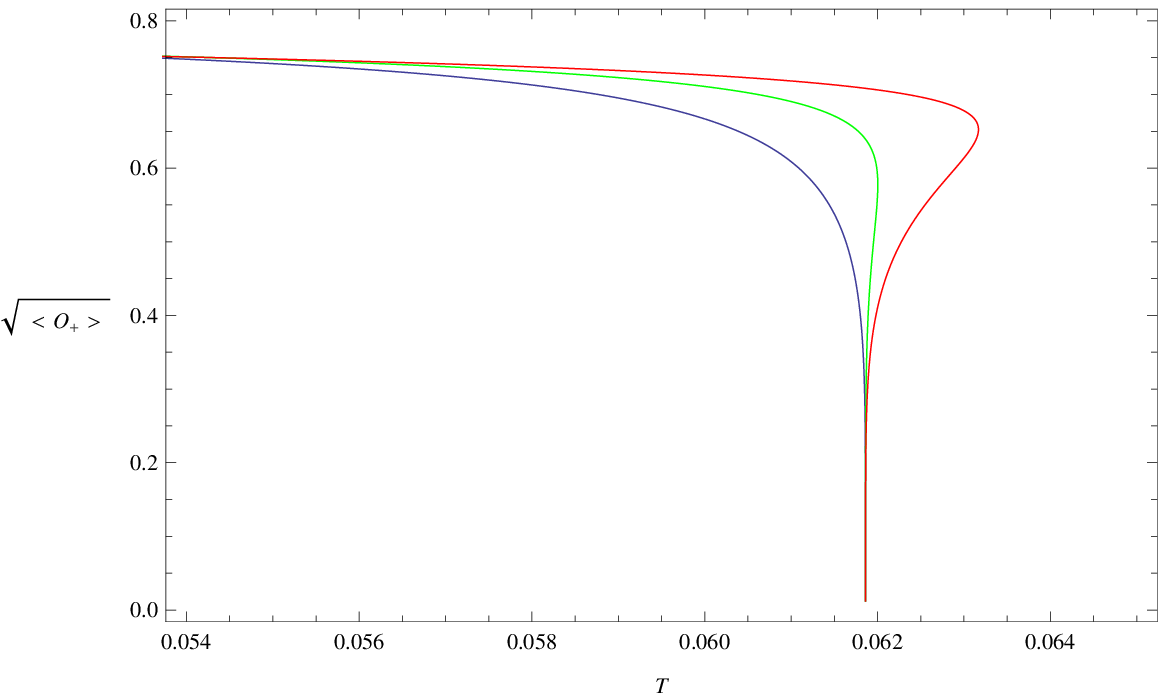}\
\caption{\label{CondBH} (Color online) The condensate $<O_{+}>$ as a
function of the temperature $T$ for fixed $\gamma=0.1$ and $\rho=1$.
The three lines in the left panel from right to left correspond to
decreasing $c_{4}$ with the fixed $m^{2}=-2$, i.e., $c_{4}=0.8$
(red), $c_{4}=0.7$ (green), $c_{4}=0.6$ (blue), the right one is for
$c_{4}=0.4$ (red), $c_{4}=0.3$ (green), $c_{4}=0.2$ (blue) with the
fixed $m^{2}=-1/2$ respectively.}
\end{figure}

\begin{figure}[h]
\includegraphics[width=220pt]{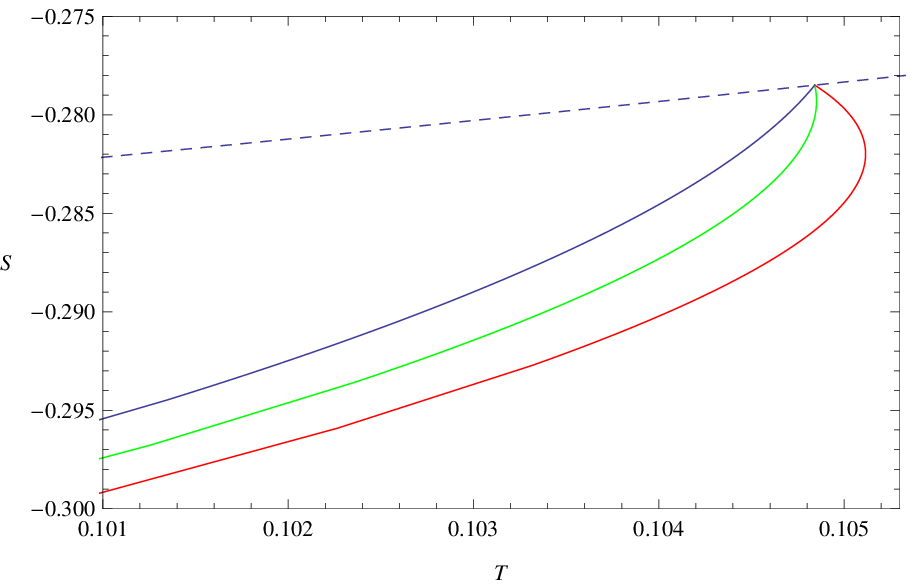}\
\includegraphics[width=220pt]{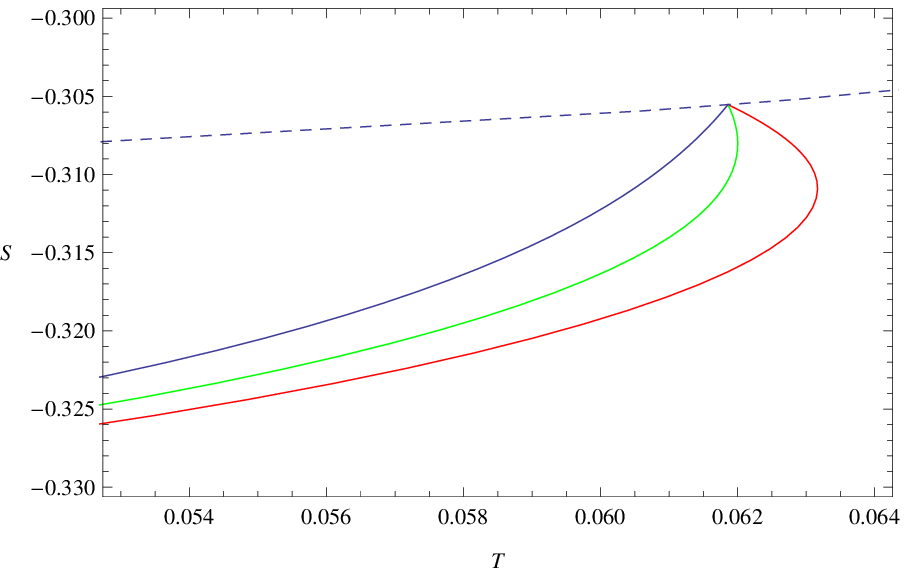}\
\caption{\label{EEntropyBHm} (Color online) The entanglement entropy
as a function of the temperature $T$ for fixed $l=1$ and
$\gamma=0.1$ with $\rho=1$. The dashed line is from the Reissner-Nordstr\"{o}m AdS
black holes and the solid curve is from the superconductor
solutions. The three solid lines in the left panel from right to
left correspond to decreasing $c_{4}$ with the fixed $m^{2}=-2$,
i.e., $c_{4}=0.8$ (red), $c_{4}=0.7$ (green), $c_{4}=0.6$ (blue),
the right one is for $c_{4}=0.4$ (red), $c_{4}=0.3$ (green),
$c_{4}=0.2$ (blue) with the fixed $m^{2}=-1/2$ respectively.}
\end{figure}

Now we want to exhibit the behavior of the entanglement entropy if
the first order phase transition appears. Generally speaking, the
order of the phase transition strongly depends on the choice of
coupling. Thus, in Fig. \ref{CondBH} we plot the condensate
$<O_{+}>$ as a function of the temperature $T$ with fixed $\rho=1$ and
$\gamma=0.1$ for different model parameters $c_{4}$. We observe
that, in this St\"{u}ckelberg model, the high correction of the
scalar field $\psi^{4}$ causes the first order phase transition for
different values of $m^{2}$. We can easily obtain the threshold
value $\overline{c}_{4}=0.7$ for the fixed $m^{2}=-2$ and
$\overline{c}_{4}=0.3$ for the fixed $m^{2}=-1/2$. Above this
threshold value, the the condensate operator does not have a
monotonic behavior, which indicates the appearance of first order
phase transition. Correspondingly, we find that in Fig.
\ref{EEntropyBHm}, where we show the entanglement entropy as a
function of the temperature $T$ for fixed $\rho=1$ and $\gamma=0.1$,
the entropy becomes multivalued near the critical temperature
$T_{c}$ and there is a discontinuous jump in the entropy if
${c}_{4}>\overline{c}_{4}$. This means that, similar to the findings
in \cite{cai-2}, the entropy can distinguish the order of phase
transition in our general superconductor model. It is interesting to
note that the jump of the entanglement entropy may be a quite
general feature for the first order phase transition.

\begin{figure}[h]
\includegraphics[width=260pt]{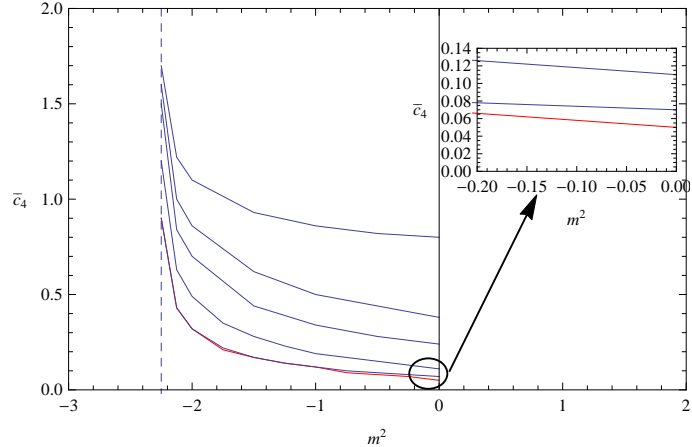}\
\caption{\label{ThresholdValue} (Color online) The threshold value
$\overline{c}_{4}$ as a function of the scalar mass $m^{2}$ for
different backreactions $\gamma$. The lines from top to bottom
correspond to the values of $\gamma$, i.e.,
$\gamma=0,~0.05,~0.10,~0.20,~0.70$ and $0.57$ (red). The vertical
dashed line is for the case $m^{2}=m_{BF}^{2}=-9/4$. Note that the
lines correspond to $\gamma=0.57$ and $\gamma=0.7$ almost coincide
with each other.}
\end{figure}

From above discussion, we note that the entanglement entropy can be
used to determine the threshold value $\overline{c}_{4}$. Thus, in
order to see the effects of the backreaction $\gamma$, mass $m$ on
$\overline{c}_{4}$ more clearly, we plot $\overline{c}_{4}$ as a
function of the scalar mass $m^{2}$ for different backreactions
$\gamma$ in Fig. \ref{ThresholdValue} by calculating the
entanglement entropy of the system. It is found that for each fixed
$\gamma$, $\overline{c}_{4}$ decreases as we increase $m^{2}$. That
is to say that the more negative mass will depress the first order
phase transition. On the other hand, for the chosen $m^{2}$, it is
shown that $\overline{c}_{4}$ decreases as we increase $\gamma$ in
the range $[0,0.57]$, but it increases very slightly as we increase
$\gamma$ when $\gamma>0.57$. Our more precise calculation shows that
$\overline{c}_{4}\rightarrow 1.74$ when $\gamma=0$ and
$m^{2}\rightarrow -9/4$, and $\overline{c}_{4}=0.05$ if
$\gamma=0.57$ and $m^{2}=0$, which means that there are an upper
limit of this threshold $\overline{c}_{4}=1.74$ and a bottom bound
$\overline{c}_{4}=0.05$. Above this upper limit, there is only the
first order phase transition for all choice of $m^{2}$ satisfying
the BF bound. If $c_{4}\in[0.05,1.74]$, we observe that $c_{4}$
combined with $m^{2}$ and $\gamma$ can trigger the first order phase
transition. Below the bottom bound, there is always the second order
phase transition, which indicates that we can not rely on the
backreaction coupled with the scalar mass to trigger the first order
phase transition with $c_{4}<0.05$. This is totally different from
the insulator/superconductor transition model in the AdS soliton
where the strong backreaction can always trigger the first order
phase transition \cite{P. Yan}. Obviously, the entanglement entropy
is powerful to explore the property of the holographic dual models.

\section{Conclusions and discussions}

We have introduced a general class of gravity dual models via
St\"{u}ckelberg mechanism and investigated the behavior of the
entanglement entropy of the systems both in the backgrounds of the
AdS soliton and AdS black hole. We noted that the holographic
entanglement entropy is a good probe to explore the properties of
the phase transition. In the AdS soliton background, by calculating
the holographic entanglement entropy for a half space in the
insulator/superconductor transition, we found that the larger
critical chemical potential corresponds to the larger maximum of the
entropy after the scalar field condensates. Furthermore, we observed
that the backreaction coupled with the scalar mass and the model
parameter can determine the order of phase transition and the more
negative mass will make the first order phase transition harder to
happen. Extending our calculation into the AdS black hole
background, we obtained the effects of the backreaction, the scalar
mass and the model parameter on the holographic entanglement entropy
for a strip shape. If the model parameter ${c}_{4}$ larger than some
threshold value determined by the backreaction and the scalar mass,
we saw that the entropy becomes multivalued near the critical
temperature and there is a discontinuous jump in the entropy, which
indicates the appearance of first order phase transition. We argued
that the jump of the entanglement entropy may be a quite general
feature for the first order phase transition. It is also interesting
to note that we can not rely on the backreaction coupled with the
scalar mass to trigger the first order phase transition if the model
parameter is below its bottom bound, which is totally different from
the insulator/superconductor transition model in the AdS soliton
where the strong backreaction can always trigger the first order
phase transition.

\begin{acknowledgments}

We thank Professor Bin Wang for his helpful discussions and
suggestions. This work was supported by the National Natural Science
Foundation of China under Grant Nos. 11305097 and 11275066; the
education department of Shaanxi province of China under Grant No.
2013JK0616; Hunan Provincial Natural Science Foundation of China
under Grant No. 12JJ4007.

\end{acknowledgments}

\end{document}